  \documentclass[preprint,12pt]{elsarticle}
 
\makeatletter
\def\ps@pprintTitle{%
 \let\@oddhead\@empty
 \let\@evenhead\@empty
 \def\@oddfoot{\centerline{\thepage}}%
 \let\@evenfoot\@oddfoot}
\makeatother

\newdefinition{exa}{Example}
\usepackage[margin=1.5in]{geometry}    % For margin alignment

\usepackage[english]{babel}
\usepackage[utf8]{inputenc}
\usepackage{algorithm}
\usepackage[noend]{algpseudocode}

\usepackage{adjustbox}
\usepackage{mathtools}
\usepackage{amssymb}
\usepackage{amsthm}

\begin{document}

\begin{frontmatter}


%% \title{Direct Nonparametric Predictive Inference Classification Trees}

\title{Direct Nonparametric Predictive Inference Classification Trees}

%\author[a,b]{Abdulmajeed Atiah Alharbi\corref{mycorrespondingauthor}}
%\ead{aahharbi@taibahu.edu.sa}
%\author[a]{Frank P.A. Coolen}
%\ead{frank.coolen@durham.ac.uk}
%\author[a]{Tahani Coolen-Maturi}
%\ead{tahani.maturi@durham.ac.uk}

%\affiliation[a]{organization={Department of Mathematical Sciences},
   %        addressline={Durham University},
     %      city={Durham},
     %      country={UK}}

%\affiliation[b]{organization={Department of Mathematics},
%	addressline={Taibah University},
%	city={Madinah},
%	country={Saudi Arabia}}

\author[a,b]{Abdulmajeed Atiah Alharbi\corref{cc}}
\ead{aahharbi@taibahu.edu.sa}
\author[a]{Frank P.A.\ Coolen}
\ead{frank.coolen@durham.ac.uk}
\author[a]{Tahani Coolen-Maturi}
\ead{tahani.maturi@durham.ac.uk}
\address[a]{Department of Mathematical Sciences, Durham University, Durham, UK.}
\address[b]{Department of Mathematics,Taibah University, Madinah, Saudi Arabia.}
\cortext[cc]{Corresponding author}

\begin{abstract}
Classification is the task of assigning a new instance to one of a set of predefined categories based on the attributes of the instance. A classification tree is one of the most commonly used techniques in the area of classification. In this paper, we introduce a novel classification tree algorithm which we call Direct Nonparametric Predictive Inference (D-NPI) classification algorithm. The D-NPI algorithm is completely based on the Nonparametric Predictive Inference (NPI) approach, and it does not use any other assumption or information. The NPI is a statistical methodology which learns from data in the absence of prior knowledge and uses only few modelling assumptions, enabled by the use of lower and upper probabilities to quantify uncertainty. Due to the predictive nature of NPI, it is well suited for classification, as the nature of classification is explicitly predictive as well. The D-NPI algorithm uses a new split criterion called Correct Indication (CI). The CI is about the informativity that the attribute variables will indicate, hence, if the attribute is very informative, it gives high lower and upper probabilities for CI. The CI reports the strength of the evidence that the attribute variables will indicate, based on the data. The CI is completely based on the NPI, and it does not use any additional concepts such as entropy. The performance of the D-NPI classification algorithm is tested against several classification algorithms using classification accuracy, in-sample accuracy and tree size on different datasets from the UCI machine learning repository. The experimental results indicate that the D-NPI classification algorithm performs well in terms of classification accuracy and in-sample accuracy.\\
\end{abstract}


\end{frontmatter}


\section{Introduction}\label{intro}
Classification is one of the most common data mining techniques that is used for assigning a new instance to one of a set of predefined categories based on the attributes of the instance. The aim of classification is to predict the unknown class states of instances which attribute values are known. There are many classification methods available in the literature, the classification tree is one of the most commonly used because of its interpretational simplicity. There are a number of algorithms that is used to build classification trees. The C4.5 algorithm \cite{quinlan93} is one of the most commonly used classical algorithms. It is an extension version of the ID3 algorithm \cite{quinlan}.\\

In recent years, theories of imprecise probabilities have been widely developed for several areas of statistics. Augustin et al.\ \cite{augustin2014introduction} have presented an overview of the main aspects of imprecise probability and its applications. Walley \cite{walley1996} has introduced the Imprecise Dirichlet Model (IDM) for inference based on multinomial data. The IDM has been used in several statistical problems \cite{bernard2005introduction}, including classification \cite{abellan2003building}. However, the use of the IDM has been criticised for some drawbacks \cite{piatti2005limits}. Thus, an alternative approach for inference from multinomial data has been presented by Coolen and Augustin \cite{unknown-c}, which is Nonparametric Predictive Inference for Multinomial data (NPI-M).\\

Nonparametric Predictive Inference (NPI) is a statistical methodology which uses only few modelling assumptions, enabled by the use of lower and upper probabilities to quantify uncertainty. NPI has been developed in recent years for different applications in statistics, operations research, risk and reliability \cite{coolen2011npi}. NPI has been introduced for several types of datasets, such as Bernoulli data \cite{Coolen_low.str}, real-valued data \cite{npi-interv-2004}, right-censored data \cite{coolen-right-cen}, ordinal data \cite{ordinal-12} and multinomial data \cite{rebecca-thesis,unknown-c,coolen-2009}. NPI is based on Hill’s assumption $ A_{(n)} $ \cite{Hills-An}, which is used for prediction about future instances with real-valued data. The $ A_{(n)} $ assumption is not suitable for multinomial data, hence, a variation of Hill’s assumption $ A_{(n)} $, which is called \textit{circular}-$A_{(n)}$ assumption, is used for multinomial data \cite{coolen2006nonparametric}.\\

In this paper, we propose a new algorithm to build classification trees using imprecise probabilities and based on the NPI approach, which we call Direct Nonparametric Predictive Inference (D-NPI) classification tree algorithm. As a first step, we introduce the D-NPI classification algorithm for binary data. After that the D-NPI classification algorithm for multinomial data is presented. The D-NPI classification algorithm can base classification on the lower and upper probabilities for events containing binary or multinomial data, without adding any further assumptions or information. The variable selection process for this algorithm, i.e.\ split criterion, is a new split method which is completely based on the NPI lower and upper probabilities. In particular, we use the NPI approach for binary and multinomial data. The split criterion is called Correct Indication ($ CI $). It does not use any additional concepts such as entropy. The $ CI $ is simply about the indication that the attribute variables will give. After computing the lower and upper probabilities of $ CI $ for all attribute variables, we aim at maximum probability for both the lower and upper probabilities of $ CI $. The D-NPI classification algorithm build classification trees using the best attribute variable selected by the $ CI $ split criterion, and stop splitting when there is no attribute variable that gives more lower and upper probabilities for $ CI $ than the lower and upper probabilities given by the NPI approach for the best class state.\\

We have carried out an experimental analysis in order to assess the performance of the D-NPI classification algorithm when building classification trees. We have also compared the performance of the D-NPI classification algorithm with other classical algorithm such as the C4.5 algorithm \cite{quinlan93}, and other imprecise algorithms based on the IDM or NPI-M such as the NPI-M \textit{algorithm} \cite{abellan2014classification,rebecca-thesis}, the A-NPI-M  \textit{algorithm} \cite{abellan2014classification,rebecca-thesis} and the IDM \textit{algorithm} \cite{abellan2003building} with two choices of the parameter $ s $. We have evaluated the performance of the classification algorithms using the classification accuracy (on testing sets), in-sample accuracy (on training sets) and average tree size. A 10-fold cross validation scheme has been applied on different datasets from the UCI repository of machine learning database \cite{uci}. The experimental results suggest that the D-NPI classification algorithm slightly outperforms the other classification algorithms in terms of both classification accuracy and in-sample accuracy. The results have also indicated that the D-NPI classification algorithm produces relatively smaller trees than the ones produced by imprecise algorithms, but slightly larger trees than the ones generated by the C4.5 algorithm.\\

The rest of the paper is organized as follows: Section \ref{backgr.} briefly describes some works related to classification trees with classic or imprecise split criteria and a brief overview of imprecise probability and NPI approach. Section \ref{D-NPI sec.} introduces the direct classification using NPI for binary or multinomial data. Section \ref{CI} presents our new split criterion, Correct Indication, for both binary and multinomial data. In Section \ref{D-NPI algorithm}, the proposed D-NPI classification algorithm is explained. Section \ref{exp. ana} describes the experimental analysis carried out on different datasets to assess and compare the proposed algorithm, with some comments on the results. Finally, conclusions and future works are given in Section \ref{concl.}.
%%%%%%%%%%%%%%%%%%%%%%%%%%%%%%%%%%%%%%%%
%%%%%%%%%%%%%%%%%%%%%%%%%%%%%%%%%%%%%%%%
\section{Background}\label{backgr.}
In this section, we review some of the most commonly used classic and imprecise split criteria that are used to build classification trees. Then, an introduction to imprecise probability is given. Finally, Nonparametric Predictive Inference (NPI) method is introduced, particularly, for binary and multinomial data.\\  

%%%%%%%%%%%%%%%%%%%%%%%%%%%%%%%%%%%%%%%%
\subsection{Classification trees}
A classification tree is a nonparametric technique which is represented by a hierarchical classifier that is based on a recursive partition method. The use of a classification tree is to classify a new observation into one of a predefined set of classes based on its attribute's values. Classification trees are mainly used on a dataset that contains one or more attribute variables and a categorical target variable. In a classification tree, each non-leaf node represents an attribute variable, each branch denotes an outcome of the attribute variable and each terminal or leaf node is assigned to one class label. Classifying a new instance is straightforward once a classification tree has been built. So, instances are classified by navigating them from the root node of the tree and going down to a terminal (leaf) node according to the outcome of the attribute variables along the path \cite{Rokach}.\\

\subsubsection{Split criteria}
A classification tree algorithm requires a split criterion which is used to select the best attribute variable to split on at each step of building classification trees. Thus, several classification tree algorithms have been developed using different split criteria. Split criteria are mainly used in order to reduce the impurity of a node. We briefly review two of the most commonly used split criteria which are Information Gain \cite{quinlan} and Information Gain Ratio \cite{quinlan93}. These split criteria are used to implement the ID3 and the C4.5 algorithms, respectively. In addition, we briefly review an imprecise split criterion called Imprecise Information Gain that has been used to construct credal classification trees \cite{abellan2003building}. These split criteria are introduced to compare some classification algorithms, which are based on them, with our new algorithm, which is based on a different split criterion. 

\paragraph{Information Gain}
The Information Gain was introduced by Quinlan \cite{quinlan} as a split criterion for the ID3 algorithm. Information Gain is an impurity-based method which uses entropy as an impurity measure. Entropy, also called the Shannon Entropy \cite{shannon1948}, of a training set $ S $ is given by

\begin{equation}\label{}
	H(S)= - \sum_{i=1}^{K} p_i \log_2 (p_i)
\end{equation}  
where $ p_i $ is the proportion of $ S $ belonging to class $ i $ (for $ i=1, ... , K $), and $ \log_2  $ is used because the information is coded in bits \cite{changala2012}. Generally speaking, entropy represents a level of uncertainty or impurity in a set of examples. The Information Gain of an attribute $ B $, relative to the training set $ S $ is given by 

\begin{equation}\label{IG}
	{Gain}(S,B)= H(S)- \sum_{u\in V(B)} \frac{|S_u|}{|S|} H(S_u)
\end{equation}
where $ V(B) $ denote all values of attribute $ B $, and $ S_u $ is the subset of $ S $ for which attribute $ B $ has value $ u $. The Information Gain handles only discrete attributes. 

\paragraph{Gain Ratio}
The Gain Ratio was introduced by Quinlan \cite{quinlan93} as an extension to the Information Gain split criterion. It is used as a split criterion for the C4.5 algorithm, hence, the C4.5 algorithm is an improved version of the ID3 algorithm. Unlike the ID3, the C4.5 algorithm handles both discrete and continuous attributes. The Information Gain is biased to attribute variables that have many states \cite{quinlan}. Thus, the Gain Ratio normalizes the Information Gain as follows:

\begin{equation}\label{}
	GR (S, B)= \frac{{Gain} (S, B)}{{SI} (S, B)}\;,
\end{equation}
where $ {Gain} (S, B) $ is given by Equation (\ref{IG}), and Split Information $ {SI} (S, B) $ is given by:

\begin{equation}\label{}
	{SI} (S, B) = - \sum_{j=1}^{n} \frac{|S_j|}{|S|} \log_2 \frac{|S_j|}{|S|}.
\end{equation}
The $ SI(S, B) $ represents the information generated by splitting the training data set $ S $ into $ n $ partitions corresponding to the outcomes of the attribute variable $ B $.

\paragraph{Imprecise Information Gain}
The Imprecise Information Gain (IIG) was introduced by Abell{\'a}n and Moral \cite{abellan2003building} to build classification trees from imprecise probability perspective. The IIG for an attribute variable $ X $ is defined as follows:
\begin{equation}
	{IIG}(X,C)= S(K(C))- \sum p(x_{i}) S(K(C|(X = x_{i}))),
\end{equation}
where $ S (K) $ is the maximum entropy of a credal set, and $ K(C) $ and $ K(C|(X = x_{i})) $ are credal sets for the class variable $ C $ and attribute variable $ C|(X = x_{i}) $, respectively; and $ p(x_{i}) $ is a probability distribution that belongs to the credal set $ K(X) $.  Credal sets are closed and convex sets of probability distributions. The IIG is applied on credal sets using uncertainty measures of probability distributions \cite{abellan2010ensemble}. For more details and extended explanations of the IIG see \cite{abellan2010ensemble,abellan2003building,abellan2005upper}. Different classification trees can be built using the IIG split criterion. For example, one can build a classification tree using the maximum entropy distributions from the credal set of distributions associated with the Imprecise Dirichlet Model (IDM) \cite{abellan2006uncertainty} or with the Nonparametric Predictive Inference for multinomial data (NPI-M) \cite{abellan2011maximising}, which are introduced in Sections \ref{imp-prob.} and \ref{npi:intro}, respectively. In this paper, we refer to a classification tree built with the IIG and the IDM by IDM \textit{algorithm}; the IIG and NPI-M by NPI-M \textit{algorithm}; and the IIG and the A-NPI-M by A-NPI-M \textit{algorithm}.\\

%%%%%%%%%%%%%%%%%%%%%%%%%%%%%%%%%%%%%%%%
\subsection{Imprecise probabilities}\label{imp-prob.}
In the middle of the 19th century, the idea of imprecise probabilities was first proposed in Boole's book \cite{boole1854investigation}. Since then, imprecise probability based methods have been developed for many areas of statistics. An overview of the main aspects of imprecise probabilities theory and applications has been presented by Augustin et al.\ \cite{augustin2014introduction}.\\

In classical probability theory, for an event $ A $, a precise probability $ p(A) \in [0, 1] $ is used to quantify uncertainty about $ A $, where $ p $ is a probability satisfying Kolmogorov’s axioms \cite{augustin2014introduction}. In real world datasets, no dataset can always indicate a precise probability, hence, having the possibility to use imprecise probability may give advantages over the use of precise probability. Imprecise probability uses lower and upper probabilities for the event, and hence reflects more uncertainty about the event. Unlike classical probability, in imprecise probability we assign an interval probability for an event $ A $, such $ [\underline{P}(A), \overline{P}(A)] $, where $ 0\leq \underline{P}(A)\leq \overline{P}(A)\leq 1 $, and where $ \underline{P}(A) $ denotes the lower probability and $ \overline{P}(A) $ denotes the upper probability for event $ A $. The classical probability is a special case in imprecise probability which occurs when $ \underline{P}(A) = \overline{P}(A) $. Complete lack of information about an event $ A $ is represented by $ \underline{P}(A) = 0 $ and $ \overline{P}(A) = 1 $. Weichselberger \cite{weichselberger2000theory} defined the structure, $ \mathcal{M} $:
\begin{equation}\label{}
	\mathcal{M}=\{p(.): \underline{P}(A) \leq p(A) \leq \overline{ P }(A), \forall A \in \mathcal{A}\}
\end{equation}
where $ \mathcal{A} $ is a set of events, and $ p(.) $ is a set-function defined on $ \mathcal{A} $ satisfying Kolmogorov’s axioms in classical probability theory. The lower and upper probabilities for an event $ A $ are:
\begin{equation}\label{}
	\underline{P}(A) = \inf \limits_{p(.) \in \mathcal{M}} p(A),
\end{equation}
and
\begin{equation}\label{}
	\overline{P}(A) = \sup  \limits_{p(.) \in \mathcal{M}} p(A).
\end{equation}
Here some basic concepts of imprecise probability are briefly highlighted. For more details and a complete introduction to imprecise probability, we refer to Walley \cite{walley1991} and Augustin et al.\ \cite{augustin2014introduction}.\\

In 1996, Walley introduced an imprecise probability model for inference from multinomial data \cite{walley1996}, which is called the Imprecise Dirichlet Model (IDM). Assume that we have a dataset with $ N $ observations. Let $ X $ be a variable whose values belongs to $ \{x_{1}, ..., x_{n}\} $, and let $ n_{x_{i}} $ denotes the total number of observations in $ x_{i} $, for $ i= 1, ..., n $. The IDM-based lower and upper probabilities for the event that the next future observation, $ X_{n+1} $ will be in $ x_{i} $ are

\begin{equation}
  \underline{P}_{IDM}(X_{n+1}\in x_{i}) = \frac {n_{x_{i}}} { N + s }
\end{equation}
and
\begin{equation}
  \overline{P}_{IDM}(X_{n+1}\in x_{i}) = \frac {n_{x_{i}}+s} { N + s }
\end{equation}
where $ s $ is a given parameter which is chosen independently of the data. The value of $ s $ determines how quickly the lower and upper probabilities converge when the sample size increases \cite{walley1996}. Walley suggested to choose the value of the parameter $ s $ equal to 1 or 2 \cite{walley1996}. As shown by Abell{\'a}n \cite{abellan2006uncertainty}, the IDM gives imprecise probabilities that lead to the following (closed and convex) credal set of probability distributions,    
\begin{equation}
	L=\left\{p \mid p\left(x_{i}\right) \in\left[\frac{n_{x_{i}}}{N+s}, \frac{n_{x_{i}}+s}{N+s}\right], \quad i=1, \ldots, n, \sum_{i=1}^{n} p(x_{i})=1\right\}.
\end{equation}

The IDM has been applied to many statistical problems in the literature. Some of these applications were reviewed by Bernard \cite{bernard2005introduction}. However, the use of the IDM has been criticised for some disadvantages \cite{piatti2005limits}. Coolen and Augustin \cite{unknown-c,coolen-2009} proposed a new model for inference from multinomial data, which is Nonparametric Predictive Inference for Multinomial data (NPI-M). This model is an alternative to the IDM and is based on the NPI method. The NPI-M does not suffer from some disadvantages of the IDM. In addition, the NPI-M does not make any prior assumptions about the data. Abell{\'a}n et al.\ \cite{abellan2014classification} and Baker \cite{rebecca-thesis} show that applying the NPI-M to classification trees leads to slightly better classification accuracy than the IDM. The NPI-M is introduced in Section \ref{npi:intro}.\\

%%%%%%%%%%%%%%%%%%%%%%%%%%%%%%%%%%%%%%%%
\subsection{Nonparametric Predictive Inference (NPI)}\label{npi:intro}
Nonparametric Predictive Inference (NPI) is a statistical methodology which uses only few modelling assumptions to learn from data in the absence of prior knowledge. NPI is based on Hill’s assumption $ A_{(n)} $ \cite{Hills-An}, explained in Section \ref{An-ass}, and uses lower and upper probabilities, also known as imprecise probabilities, to quantify uncertainty \cite{npi-interv-2004}. NPI has been developed in recent years for different applications in statistics, operations research, risk and reliability \cite{coolen2011npi}. NPI has been presented for different types of data, such as Bernoulli data \cite{Coolen_low.str}, real-valued data \cite{npi-interv-2004}, right-censored data \cite{coolen-right-cen}, ordinal data \cite{ordinal-12} and multinomial data \cite{rebecca-thesis,unknown-c,coolen-2009}.\\

\subsubsection{$A_{(n)}$ assumption}\label{An-ass}
Hill \cite{Hills-An} introduced the assumption $A_{(n)}$ for prediction about future observations when there is no strong prior knowledge about the form of the underlying distribution of a random quantity.  Hill’s assumption $ A_{(n)} $ directly provides probabilities for one or more real-valued future random quantities, based on observed values of related random qualities. Let $ X_{1}, ... , X_{n}, X_{n+1} $ be real-valued and exchangeable random quantities, where we assume that the probability of ties is zero. Assume that $ X_{1}, ... , X_{n} $ be ordered and denoted as $ x_{(1)} < ... < x_{(n)} $, and let $ x_{0}= -\infty  $ and $ x_{n+1}= \infty  $ for ease of notation. These ordered observations partition the real line into $ n + 1 $
open intervals $ I_{j} = (x_{j-1}, x_{j})$ for each $ j = 1, ... , n + 1 $. The assumption $ A_{(n)} $ states that the next future observation, represented by a random quantity $ X_{n+1} $, falls equally likely in any interval $ I_j $ with probability $ \frac{1}{n+1} $ for each $ j = 1, ... , n + 1 $, i.e. $ P(X_{n+1} \in I_{j})= \frac{1}{n+1} $. Hill’s assumption $ A_{(n)} $ does not assume anything else, and it is clearly a post-data assumption related to exchangeability of $ n+1 $ values on the real-line. \cite{theoryofprb}.\\  

\subsubsection{NPI for Bernoulli data} \label{npi-bernoulli}
This section summarises NPI for Bernoulli random quantities as introduced by Coolen \cite{Coolen_low.str}. Suppose that we have a sequence of $ n + m $ exchangeable Bernoulli trials where the possible outcomes of each trial are either `success' or `failure', and the data consist of $ s $ successes in $ n $ trials. Let $ Y^{n}_{1} $ and $ Y^{n+m}_{n+1} $ denote the random number of successes in trials $ 1 $ to $ n $,  and $ n+1 $ to $ n+m $, respectively. Because of the assumed exchangeability of all trials, a sufficient representation of the data for the inference considered is $ Y^{n}_{1} = s $. Now, let $ R_{t}=\{r_{1}, ... ,r_{t}\} $, with $ 1 \leq t \leq m + 1 $ and $ 0 \leq r _ { 1 } < r _ { 2 } < \ldots < r _ { t } \leq m $, and for ease of notation, let $\big(\begin{smallmatrix}s + r _ { 0 }\\{s}\end{smallmatrix}\big) = 0$. Then, the NPI upper probability for the conditional event $ Y_{n+1}^{n+m} \in R_{t} | Y_{1}^{n}=s, \; \mbox{for} \; s \in\{0, \ldots, n\}$, is

\begin{multline}\label{upper-Berno}
	\overline{P}\left(Y_{n+1}^{n+m} \in R_{t} | Y_{1}^{n}=s\right) = \\
	\left( \begin{array}{c}{n+m} \\ {n}\end{array}\right)^{-1} \sum_{j=1}^{t}\left[\left( \begin{array}{c}{s+r_{j}} \\ {s}\end{array}\right)-\left( \begin{array}{c}{s+r_{j-1}} \\ {s}\end{array}\right)\right] \left( \begin{array}{c}{n-s+m-r_{j}} \\ {n-s}\end{array}\right).
\end{multline}\\
It is sufficient to determine the NPI upper probability only, and the corresponding NPI lower probability, can be derived by the conjugacy property, $ \underline{P}(A) = 1- \overline{ P }(A^{c}) $, where $ A^{c} $ is the complementary event to $ A $.
\begin{equation}\label{lower-Berno}
	\underline{P}\left(Y_{n+1}^{n+m} \in R_{t} | Y_{1}^{n}=s\right)=1-\overline{P}\left(Y_{n+1}^{n+m} \in R_{t}^{c} | Y_{1}^{n}=s\right)
\end{equation}
where $ R_{t}^{c}=\{0,1, \ldots, m\} \backslash R_{t} $. More details and examples about the NPI for Bernoulli quantities are give by Coolen \cite{Coolen_low.str}.\\

\subsubsection{NPI for multinomial data} \label{npi-mul}
Coolen and Augustin \cite{npi-interv-2004,unknown-c,coolen-2009} have developed Nonparametric Predictive Inference for multinomial data (NPI-M). The NPI-M is based on
\textit{circular}-$A_{(n)}$ assumption, which is related to predictive inference involving multinomial data. The $ A_{(n)} $ assumption is not suitable for multinomial data because of that multinomial data are represented as observations on a probability wheel instead of the real-line. Therefore, a variation of Hill’s assumption $ A_{(n)} $ which is called \textit{circular}-$A_{(n)}$ and denoted by $ \textcircled{\small{A}}_{(n)} $ is used \cite{coolen2006nonparametric,unknown-c}. Suppose that we have ordered circular data $ y_{1} < y_{2} < ... <y_{n} $ which create $ n $ intervals on a circle, represented as $ I_{j}= (y_{j}, y_{j+1}) $ for $ j= 1, ..., n-1 $ and $ I_{n}= (y_{n}, y_{1}) $. The assumption $ \textcircled{\small{A}}_{(n)} $ states that the next future observation, represented by a random quantity $ Y_{n+1} $, falls equally likely in any interval $ I_j $ with probability $ \frac{1}{n} $ for each $ j = 1, ... , n $, i.e. $ P(Y_{n+1} \in I_{j})= \frac{1}{n} $. The $ \textcircled{\small{A}}_{(n)} $ is a post-data assumption related to exchangeability for such circular data. The NPI-M model represents multinomial data as observations on a probability wheel, where each of these $ n $ observations is represented by a line from the center of the wheel to its circumference. Thus, the wheel is divided into $ n $ equally-sized slices. Using \textit{circular}-$ A_{(n)} $ assumption, the next future observation has probability mass $ \frac{1}{n} $ of falling into any given slice. Hence, we have to decide which category each of these slices represents.\\

Coolen and Augustin \cite{coolen-2009} assume that each observed category is represented by one single segment of the probability wheel, where the segment is an area between two lines from the center to the circumference of the wheel. Combining this assumption with \textit{circular}-$ A_{(n)} $ implies that two or more lines representing observations in the same categories are positioned next to each other. Therefore, a slice that is bordered  by two lines representing different categories is a separating slice, which could be assigned to any of these different categories or to unobserved categories. It is also assumed that there is no ordering of the categories, and hence no specific ordering of the segments on the wheel.\\

Coolen and Augustin introduced the NPI-M for the case of known number of categories \cite{coolen-2009} and for the case of unknown number of categories \cite{unknown-c}. We restrict our focus in this paper on the case where the number of possible categories, denoted by $ K $ is known. We assume that $ K\geq 3 $. However, for the case when $ K= 2 $, the NPI-M can also be used, but using NPI for Bernoulli data \cite{Coolen_low.str} is more appropriate as it leads to slightly less imprecision. In the following, we summarise the results of \cite{coolen-2009} when $ K $ is known, and similar notation to Coolen and Augustin \cite{coolen-2009} are used.\\

Suppose that there are $ K \geq 3 $ possible categories denoted by $ C_{1}, ... , C_{K} $. We assume that $ C_{1}, ... , C_{k} $ for $ 1 \leq k \leq K $ are observed categories, and $ C_{k+1}, ... , C_{K} $ are unobserved categories. Let $ n_{j} $ represents the number of observations in category $ C_{j} $ for $ j= 1, ..., k $, hence $ \sum_{j=1}^{k} n_{j} = n $. The general event of interest, can be denoted by 
\begin{equation}\label{y-n+1}
	Y_{n+1} \in \bigcup_{j \in J} C_{j}
\end{equation}
where $ J \subseteq \{1, ..., K\} $. Let $ OJ = J \bigcap \{1, ..., k\}$ and $ UJ= J \bigcap \{k+1, ..., K\} $, where $ OJ $ represent the index-set for the categories in the event of interest that have been observed, and $ UJ $ represent the index-set for the unobserved categories. Let $ r = |OJ| $ and $ l = |UJ| $, hence $ 0 \leq r \leq k $ and $ 0 \leq l \leq K - k $. This leads to $ k - r $ observed categories which are not included the event of interest, and $ K - k - l $ unobserved categories which are not included in the event of interest.\\

To find the NPI lower and upper probabilities for the event of interest, we have to consider all the possible configurations $ \sigma $ on the probability wheel. The NPI lower probability for the event of interest is achieved by selecting $ \sigma $ that minimises the number of slices assigned to the event, and the NPI upper probability for the event of interest is achieved by selecting $ \sigma $ that maximises the number of slices assigned to the event.\\

The NPI-M lower probability for the event of interest (\ref{y-n+1}) based on $ n $ observations is 
\begin{equation}\label{npi-m-l}
	\underline{P}(Y_{n+1} \in \bigcup_{j \in J} C_{j}) = \frac{1}{n}\left(\sum_{j \in OJ} n_{j} - r + \max(2r + l - K, 0)\right) 
\end{equation}
and the NPI-M upper probability for the event of interest (\ref{y-n+1}) based on $ n $ observations is 
\begin{equation}\label{npi-m-u}
	\overline{P}(Y_{n+1} \in \bigcup_{j \in J} C_{j}) = \frac{1}{n}\left(\sum_{j \in OJ} n_{j} - r + \min(2r + l, k)\right). 
\end{equation}\\

The derivations of the NPI-M lower and upper probabilities (\ref{npi-m-l}) and (\ref{npi-m-u}) are explained, with more details, examples and discussions in \cite{coolen-2009}. Coolen and Augustin \cite{coolen-2009} also presented some fundamental properties of NPI-M lower and upper probabilities.\\

For events $ Y_{n+1} \in C_{i} $, so considering only a single category, the NPI-M lower and upper probabilities are 
\begin{equation}\label{s-npi-m-l}
	\underline{P}\left(Y_{n+1} \in C_{i}\right)= \max \left(0, \frac{n_{i}-1}{n}\right) 
\end{equation}
and
\begin{equation}\label{s-npi-m-u}
	\overline{P}\left(Y_{n+1} \in C_{i}\right)=\min \left(\frac{n_{i}+1}{n}, 1\right)
\end{equation}
where $ i= 1, ..., K $ and $ n_{i} $, the number of observations in category $ i $. For classification problems, generally events with only observed categories are considered. Therefore, we consider the case that these single categories have been observed.\\
%%%%%%%%%%%%%%%%%%%%%%%%%%%%%%%%%%%%%%%%
%%%%%%%%%%%%%%%%%%%%%%%%%%%%%%%%%%%%%%%%
\section{Direct NPI classification}\label{D-NPI sec.}
In this section, we consider direct classification using NPI for binary data and NPI for multinomial data. The direct NPI classification can base classification on the lower and upper probabilities for events with binary or multinomial data, and without adding any further assumptions or information. We will first introduce direct NPI classification for binary data in Section \ref{dnpi-B}, then direct NPI classification for multinomial data is introduced in Section \ref{dnpi-M}.

%%%%%%%%%%%%%%%%%%%%%%%%%%%%%%%%%%%%%%%%
\subsection{Direct NPI classification for binary data}\label{dnpi-B}
In this section we introduce direct NPI classification for binary data. As a first step to develop the method of direct NPI classification, we start with exploring the method on complete binary data, where both the class variable and the attribute variables are binary. Assume that we have a data set of $ n $ instances which only have two outcomes, $ 0 $ or $ 1 $. These instances can also be indicated as `negative' or `positive' cases. Suppose that there are $ T\geq1 $ binary attribute variables. Let $ t_{j} $ for $ j\in\{1, ..., T\} $ indicate attribute variables. The result of each attribute is either  $ 0 $ or $ 1 $. Let $ D $ be a binary class variable, where $ D = 0 $ or $ D = 1 $. Let $ n^0 $ denote the total number of instances with $ D = 0 $, and let $ n^1 $ denote the total number of instances with $ D = 1 $, so $ n= n^{0} + n^{1} $.\\

Now, we will see if attribute $ j $ of the $ T $ available attributes is useful or not. The attribute $ j $, indicated by $ t_{j} $ is useful if an instance with attribute result $ t_{j} = 1 $, has a high probability of being classified as $ D=1 $, and an instance with attribute result $ t_{j} = 0 $, has a high probability of being classified as $ D=0 $. Therefore, we are interested in the conditional events ($ D=1 | t_{j}=1 $) and ($ D=0 | t_{j}=0 $). These conditional events indicate that attribute result $ 1 $ ($ t_{j} = 1  $) is related to class state $ 1 $ ($ D=1 $) in terms of the dataset, and similarly for ($ D=0 | t_{j}=0 $). We consider such events for one future instance for which the attributes are available but we do not know its class states. This instance is assumed to be exchangeable with all other $ n $ instances in the dataset. Let $ n(t_{j} = 1) $ be the total number of instances in the data set which have results $ t_{j} = 1 $. Let $ n^{1}(t_{j} = 1) $ denote the total number of instance which have results $ t_{j} = 1 $ and which are classified as $ D = 1 $, and $ n^{0}(t_{j} = 1) $ indicates the total number of instances which have results $ t_{j}=1 $ but are classified as $ D = 0 $. So, $ n(t_{j} = 1) = n^{1}(t_{j} = 1) + n^{0}(t_{j} = 1) $. Note that these numbers are known from the dataset.\\ 

We denote the unknown class state of one future instance which is not included in the dataset by $ D_{n+1} $. Now, using NPI for Bernoulli data \cite{Coolen_low.str}, introduced in Section \ref{backgr.}, we can derive the NPI lower and upper probabilities for $ D_{n+1} = 1 | t_{n+1,j} = 1 $. Note here that $ t_{n+1,j} = 1 $ is the attribute result for this future instance. The NPI lower probability is 

\begin{equation}\label{ld1t1}
	\underline { P } \left( D _ { n + 1 } = 1 | t _ { n+1,j } = 1 \right) = \frac { n ^ { 1 } \left( t _ { j } = 1 \right) } { n \left( t _ { j } = 1 \right) + 1 },
\end{equation}
and the NPI upper probability is

\begin{equation}\label{ud1t1}
	\overline { P } \left( D _ { n + 1 } = 1 | t _ { n+1,j } = 1 \right) = \frac { n ^ { 1 } \left( t _ { j } = 1 \right) + 1 } { n \left( t _ { j } = 1 \right) + 1 }.
\end{equation}\\
Similarly, the NPI lower and upper probabilities for $ D_{n+1} = 0 | t_{n+1,j} = 0 $ are

\begin{equation}\label{}
	\underline { P } \left( D _ { n + 1 } = 0 | t _ { n+1,j } = 0 \right) = \frac { n ^ { 0 } \left( t _ { j } = 0 \right) } { n \left( t _ { j } = 0 \right) + 1 },
\end{equation}
and
\begin{equation}\label{}
	\overline { P } \left( D _ { n + 1 } = 0 | t _ { n+1,j } = 0 \right) = \frac { n ^ { 0 } \left( t _ { j } = 0 \right) + 1 } { n \left( t _ { j } = 0 \right) + 1 }.
\end{equation}\\

The NPI lower and upper probabilities for events $ D_{n+1} = 0 | t_{n+1,j} = 1 $ and  $ D_{n+1} = 1 | t_{n+1,j} = 0 $ can be also derived via the conjugacy property. It should be noticed that the values of the above NPI lower and upper probabilities will report the strength of the evidence for the class state of the future instance, which is not included in the data, but is assumed to be exchangeable with the $ n $ other instances in the dataset. \\

%%%%%%%%%%%%%%%%%%%%%%%%%%%%%%%%%%%%%%%%
\subsection{Direct NPI classification for multinomial data}\label{dnpi-M}
In this section, we illustrate how we can base classification on the NPI-M lower and upper probabilities without adding any further assumptions or information. Assume that we have a data set of $ n $ instances. Suppose that there are $ T\geq 1 $ attribute variables, where the result of each attribute is multinomial data with known number of categories $ k $, where these categories are denoted by $ c_{1}, ..., c_{k} $. Note that in Section \ref{backgr.}, we denote the number of possible categories by $ K $ as we have $ k $ observed categories and $ k+1 $ to $ K $ unobserved categories. In our direct NPI classification method we assume all categories are observed and denoted by $ k $. Throughout this paper we refer to target classes by $ C_{i} $, and we refer to attribute's categories by $ c_{i} $. Let $ t_{j} $ for $ j=\{1, ..., T\} $ indicate attribute variables. Suppose also that we have a target variable with known number of classes. Now, we assume that the target variable is represented by class variable, $ D $, where $ D \in \{C_1, ..., C_k\} $. Let $ n^{C_{i}} $ be the number of instances which are classified as class $ i $, for $ i= 1, ..., k $, so $ n=  n^{C_{1}}+  n^{C_{2}}+ ... + n^{C_{k}} $. Let also $ n(t_{j}=c_{i}) $ be the total number of instances in the data which have $ t_{j} = c_{i} $. Now, let $ n^{C_{1}}(t_{j}=c_{1}) $ be the number of instances which have $ t_{j} = c_{1} $ and which are classified as $ C_{1} $. Thus, $ n(t_{j}=c_{1})= \sum_{i=1}^{k} n^{C_{i}}(t_{j}=c_{1}) $ denote the total number of instances in the data which have $ t_{j} = c_{1} $.\\

Now we will see if attribute $ t_{j} $ of the $ T $ attributes is useful or not. Clearly, $ t_{j} $ is useful if there is a high probability that an instance with attribute result $ t_{j}= c_1 $ indeed has class $ C_{1} $ and  an instance with attribute result $ t_{j}= c_2 $ indeed has class $ C_{2} $, and so on. Therefore, we are interested in the conditional events $ (D = C_{i}| t_{j} = c_{i}) $ for $ i=1, ..., k $. Clearly, the number of attribute outcomes $ c_{i} $ might not be the same as the number of possible states in the class variable $ C_{i} $. In this case, we consider the state of the highest frequency of possible states corresponding to each attribute outcome. In other words, each attribute category is linked with the class state that contains the largest number of instances. We consider such events for one future instance for which the attributes results are known but which class status is unknown. This instance is assumed to be exchangeable with the $ n $ instances in the data. Let $ D_{n+1} $ denote the unknown class status for a single future instance which is not included in the data. Using NPI for Bernoulli data \cite{Coolen_low.str}, introduced in Section \ref{backgr.}, we can provide the NPI lower and upper probabilities for the event that a future instance, which is not included in the data has class $ i $, $ C_{i} $ given that its attribute result is $ c_{i} $, $ t_{n+1, j} = c_{i} $, for $ i= 1, ..., k $ i.e.  $ D_{n+1} = C_{i}| t_{n+1, j} = c_{i}$, for $ i= 1, ..., k $. The NPI lower probability is

\begin{equation}\label{lnpi-b}
	\underline { P } \left( D _ { n + 1 } = C_{i} | t _ { n+1,j } = c_{i} \right) = \frac { n ^ { C_{i} } \left( t _ { j } = c_{i} \right) } { n \left( t _ { j } = c_{i} \right) + 1 },
\end{equation}
and the NPI upper probability is
\begin{equation}\label{unpi-b}
	\overline { P } \left( D _ { n + 1 } = C_{i} | t _ { n+1,j } = c_{i} \right) = \frac { n ^ { C_{i} } \left( t _ { j } = c_{i} \right) + 1 } { n \left( t _ { j } = c_{i} \right) + 1 }.
\end{equation}\\
The direct NPI classification for the conditional event $  D _ { n + 1 } = C_{i} | t _ { n+1,j } = c_{i} $ can be calculated via Formulas (\ref{lnpi-b}) and (\ref{unpi-b}). It should be noticed that the values of these NPI lower and upper probabilities will directly report the strength of the evidence for which of class states the single future instance will have, based on the data.\\
%%%%%%%%%%%%%%%%%%%%%%%%%%%%%%%%%%%%%%%%
%%%%%%%%%%%%%%%%%%%%%%%%%%%%%%%%%%%%%%%%
\section{Correct Indication}\label{CI}
In this section we introduce a novel split criterion to be used by the D-NPI algorithm when building a classification tree. Our split criterion is called \textit{Correct Indication} ($ CI $). Generally, the most informative attribute is desired in a classification tree. In the $ CI $ split criterion, if the attribute is very informative in all cases, then it gives high lower and upper probabilities for $ CI $. Thus, we aim at the maximum values for both lower and upper probabilities for $ CI $. 

%%%%%%%%%%%%%%%%%%%%%%%%%%%%%%%%%%%%%%%%
\subsection{Correct Indication for binary data}\label{CI:B}
In this section, we introduce the lower and upper probabilities of $ CI $ corresponding to binary data. Let $ p = P(t_{j} = 1) $, hence, $ P(t_{j} = 0) = 1-p $. Using NPI for Bernoulli quantities \cite{Coolen_low.str}, introduced in Section \ref{backgr.}, we get

\begin{equation}\label{p-binary}
	p \in \left[\frac {n( t _ { j } = 1)} { n + 1 }, \frac {n( t _ { j } = 1)+1} { n + 1 }\right].
\end{equation}\\
Now we can determine the NPI lower probability for attribute $ t_{j} $ to lead to $ CI $ by taking the NPI lower probabilities for $ (D_{n+1}=0|t_{n+1,j} = 0) $, and for $ (D_{n+1}=1|t_{n+1,j} = 1) $, and $ p $ within the range given by (\ref{p-binary}) to minimise the weighted average, 

\begin{equation}\label{LCI}
	\underline { P } _ { j } ( C I ) = \min\limits_p \left( \frac { n ^ { 0 } \left( t _ { j } = 0 \right) } { n \left( t _ { j } = 0 \right) + 1 } ( 1 - p ) + \frac { n ^ { 1 } \left( t _ { j } = 1 \right) } { n \left( t _ { j } = 1 \right) + 1 } p \right),
\end{equation}
Hence, we set $ p $ such that

\begin{equation}
	p = \left\{ \begin{array}{ll}
		\frac{n(t_{j} =1) + 1}{n + 1}   & \text{if} \;\;\; \frac{n^{0}(t_{j} =0)}{n(t_{j} =0) + 1} \geq \frac{n^{1}(t_{j} =1)}{n(t_{j} =1) + 1}, \\
		\frac{n(t_{j} =1)}{n + 1}  & \text{otherwise}.
	\end{array} \right.
\end{equation}\\
Similarly, the NPI upper probability for attribute $ t_{j} $ to lead to $ CI $ is given by

\begin{equation}\label{UCI}
	\overline { P } _ { j } ( C I ) = \max\limits_p \left( \frac { n ^ { 0 } \left( t _ { j } = 0 \right) + 1 } { n \left( t _ { j } = 0 \right) + 1 } ( 1 - p ) + \frac { n ^ { 1 } \left( t _ { j } = 1 \right) + 1 } { n \left( t _ { j } = 1 \right) + 1 } p \right),
\end{equation}
where $ p $ is such that

\begin{equation}
	p = \left\{ \begin{array}{ll}
		\frac{n(t_{j} =1) + 1}{n + 1}   &  \text{if} \;\;\; \frac{n^{0}(t_{j} =0)+1}{n(t_{j} =0) + 1} \leq \frac{n^{1}(t_{j} =1)+1}{n(t_{j} =1) + 1}, \\
		\frac{n(t_{j} =1)}{n + 1}  & \text{otherwise}.
	\end{array} \right.
\end{equation}\\

We aim at the maximum probability for $ CI $ for both lower and upper probabilities. Generally, in a classification tree, the most informative attribute variable is desired. In $ CI $ for binary data, if the attribute variable is very informative in both cases, then it gives high lower and upper probabilities of $ CI $.\\

%%%%%%%%%%%%%%%%%%%%%%%%%%%%%%%%%%%%%%%%
\subsection{Correct Indication for multinomial data}\label{CI:M}
In this paper, we assume that we have a known number of categories, $ k \geq 3 $. Recall that $ k $ is the number of observed categories.\\

Let $ p_{i} = P(t_{j} = c_{i}) $ for $ i = 1, ..., k $, where $ \sum_{i=1}^{k} p_{i} = 1 $. Using NPI for multinomial data \cite{coolen-2009}, introduced in Section \ref{backgr.}, we get
\begin{equation}\label{p-mult}
	p_{i} \in \left[\frac{n_{i} - 1}{n}, \frac{n_{i} + 1}{n}\right]
\end{equation}
where $ n_{i} $ denotes the number of times we have observed categories $ c_{1}, ..., c_{k} $ respectively, and $ n_{i} > 0 $.\\

Now we can determine the NPI lower probability for attribute $ t_{j} $ to lead to $ CI $, by taking the NPI lower probabilities for the events $ D_{n+1} = C_{1}| t_{n+1, j} = c_{1}, ..., D_{n+1} = C_{k}| t_{n+1, j} = c_{k} $, and $ p_{i} $ within the range given by (\ref{p-mult}) to minimise the weighted average,

\begin{equation}\label{l-ci}
	\underline { P } _ { j } ( C I ) = \min\limits_{p_{i} \in \mathcal{P}} \sum_{i=1}^{k} \frac { n ^ { C_{i} } \left( t _ { j } = c_{i} \right) } { n \left( t _ { j } = c_{i} \right) + 1 } \; p_{i}.
\end{equation}\\
Similarly, the NPI upper probability for attribute $ t_{j} $ to lead to $ CI $ is

\begin{equation}\label{u-ci}
	\overline { P } _ { j } ( C I ) = \max\limits_{p_{i} \in \mathcal{P}} \sum_{i=1}^{k} \frac { n ^ { C_{i} } \left( t _ { j } = c_{i} \right) + 1} { n \left( t _ { j } = c_{i} \right) + 1 } \; p_{i}.
\end{equation}\\
where $\mathcal{P} $ is a structure of probability distribution defined as follows:
\begin{equation}
	\mathcal{P} = \left\lbrace p \; | \; \frac{n_{i} - 1}{n} \leq p_i \leq \frac{n_{i} + 1}{n}, \forall i=1, ..., k, \sum_{i=1}^{k} p_{i} = 1\right\rbrace. 
\end{equation}\\
The above lower and upper probabilities for $ CI $ can be calculated for each single attribute, then we choose the most informative attribute at each stage of building classification tree based on these probabilities.\\

To compute the above NPI lower and upper probabilities for $ CI $, we consider all possible configurations $ \sigma $ on the probability wheel (see Section \ref{backgr.}), applying \textit{circular}-$A_{(n)}$ assumption to each $ \sigma $ to get corresponding lower and upper probabilities $ \underline{P}_{\sigma}(CI) $ and $ \overline { P } _ { \sigma } ( C I ) $ and then we take the lower and upper envelope with respect to the set $ \sum $ of all configurations $ \sigma $ such that 

\begin{equation}
	\underline{P}(CI) = \min \limits_{\sigma \in \sum} \underline{P}_{\sigma}(CI)
\end{equation}
and
\begin{equation}
	\overline{P}(CI) = \max \limits_{\sigma \in \sum} \overline{P}_{\sigma}(CI).
\end{equation} Thus, we need to optimise $ p_{i} $ over all possible configurations on the probability wheel, then to choose the configuration that gives the optimal value for $ CI $, either minimal for $ \underline { P } _ { j } ( C I ) $ or maximal for $ \overline { P } _ { j } ( C I ) $. For a large number of categories, consequently, a large number of possible configurations, we may need to fix a certain configuration to obtain the NPI lower and upper probabilities for $ CI $. Therefore, we clarify in this section which configuration is optimal to achieve the NPI lower and upper probabilities for $ CI $.\\   

We now explain how to do this optimisation by considering the NPI lower and upper probabilities separately. We first consider the NPI lower probability for $ CI $, Formula (\ref{l-ci}), and separate two cases. First, if the number of possible categories $ k $ is even. Secondly, if $ k $ is odd. After that we consider the same two cases with regards to the NPI upper probability for $ CI $, Formula (\ref{u-ci}).

\subsubsection{Lower probability}
To find the lower probability of $ CI $, $ \underline { P } _ { j } ( C I ) $, we need to minimise over all possible configurations of the probability wheel, then to choose the configuration that gives the smallest possible value.\\

Let $ f_{i} = \displaystyle \frac { n ^ { C_{i} } ( t _ { j } = c_{i} ) } { n ( t _ { j } = c_{i} ) + 1 } $, for $ i= 1, ..., k $, so we rewrite the Equation (\ref{l-ci}) in the following way 
\begin{equation}\label{}
	\underline { P } _ { j } ( C I ) = \min\limits_{p_{i} \in \mathcal{P}} (f_{1} p_{1} + f_{2} p_{2} + ... + f_{k} p_{k})
\end{equation}
Suppose that the fractions $ f_{i} $ are reordered in an increasing way and relabelled such that $ \acute{f}_{1} \leq \acute{f}_{2} \leq ... \leq \acute{f}_{k} $, with corresponding $ \acute{p}_{i} $, where $ \acute{p}_{i} $ is corresponding to the reordered $ f_{i} $. For example, if the smallest $ f_{i} $ is $ f_{2} $, then $ \acute{f}_{1}= f_{2} $ and $ \acute{p}_{1}= p_{2} $. So we get 
\begin{equation}\label{}
	\underline { P } _ { j } ( C I ) = \min\limits_{\acute{p}_{i} \in \mathcal{P}} (\acute{f}_{1} \acute{p}_{1} + \acute{f}_{2} \acute{p}_{2} + ... + \acute{f}_{k} \acute{p}_{k})
\end{equation}

\begin{figure}[t]
	\centering
	\includegraphics[width=0.45\textwidth]{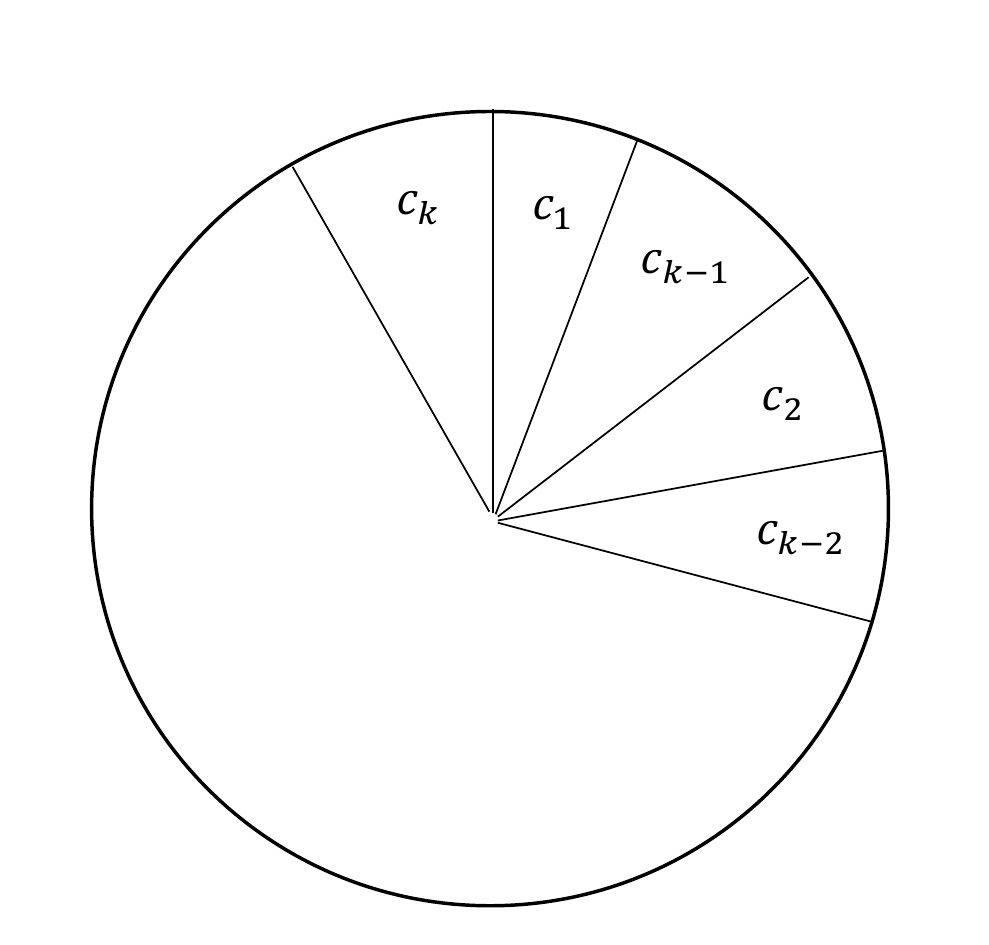}
	\caption{Optimal configuration.}
	\label{opt-conf}
\end{figure}

We separate categories corresponding to the largest $ \acute{f_{i}} $ as much as possible to ensure that we can assign probability masses of slices `in between' to its neighbour with smaller value of $ \acute{f_{i}} $, for minimisation. Therefore, the configuration of the wheel which gives the most flexibility to do so is the arrangement where categories $ c_{1}, ..., c_{k} $ corresponding to $ \acute{p}_{1}, ..., \acute{p}_{k} $ are permuted in the following way, $ c_{1}, c_{k-1}, c_{2}, c_{k-2}, c_{3}, c_{k-3}, ...., c_{k} $. This arrangement can be represented on the probability wheel as in Figure \ref{opt-conf}.\\

Generally speaking, to find the lower probability of $ CI $, each category is assigned its lower probability,  $ \frac{n_{i}- 1}{n} $ and the remaining probability mass is then shared between the categories with the smallest $ \acute{f_{i}} $ in such a way to derive the smallest value of the lower probability of $ CI $. However, the way in which this can be shared must not violate the rules of the NPI for multinomial data. Hence, only $ \frac{1}{n} $ can be assigned from either side to a category, which implies at most $ \frac{2}{n} $ in total to any category. We now consider two cases: First, when $ k $ is even. Then, when $ k $ is odd. 

\paragraph{Case 1: $ k $ is even}
To find the lower probability of $ CI $, we initially assign probability mass $ \frac{n_{i}- 1}{n} $ to each category. Once this probability assignments are made, there are $ k $ separating slices remaining. These separating slices must be then shared equally between the categories with the smallest fractions $ \acute{f_{i}} $, provided that the resulting probabilities are no larger than their upper limits $ \frac{n_{i} + 1}{n} $.  Therefore, the remaining probability masses $ \frac{k}{n} $ must be distributed equally from $ {c}_{1} $ to $ {c}_{\frac{k}{2}} $, hence, we assign additional probability mass of $ \frac{2}{n} $ to each of $ {c}_{1} $ to $ {c}_{\frac{k}{2}} $.\\

\begin{exa} \label{k6}
	Suppose we have 6 possible categories named, $ c_{1}, c_{2}, c_{3}, c_{4}, c_{5} $ and $ c_{6} $ where $ (n_{1}, n_{2}, n_{3}, n_{4}, n_{5}, n_{6}) = (1, 2, 3, 4, 5, 6)$, and hence, $ n=21 $. For simplicity, suppose also that we reorder their corresponding fractions $ f_{i} $, for $ i =1, ..., 6 $ in an increasing order such that $ \acute{f_{1}} \leq ... \leq \acute{f_{6}} $. Then the NPI lower probability for $ CI $ for attribute $ j $ is
	\begin{equation}\label{}
		\underline { P } _ { j } ( C I ) = \min\limits_{\acute{p}_{i} \in \mathcal{P}} (\acute{f}_{1} \acute{p}_{1} + \acute{f}_{2} \acute{p}_{2} + ... + \acute{f}_{6} \acute{p}_{6})
	\end{equation}
	
	Using NPI for multinomial data, each category is assigned its lower probability $ \frac{n_{i} - 1}{n} $. Thus, we assign $ (0, \frac{1}{21}, \frac{2}{21}, \frac{3}{21}, \frac{4}{21}, \frac{5}{21}) $ to $ (c_{1}, c_{2}, c_{3}, c_{4}, c_{5}, c_{6}) $, respectively. Then we assign $ (\frac{2}{21}, \frac{2}{21}, \frac{2}{21}) $ to  $ (c_{1}, c_{2}, c_{3}) $. Therefore, the final assignments are $ (\frac{2}{21}, \frac{3}{21}, \frac{4}{21}, \frac{3}{21}, \frac{4}{21}, \frac{5}{21}) $ to $ (c_{1}, c_{2}, c_{3}, c_{4}, c_{5}, c_{6}) $, respectively. Following theses assignments we get $ \underline { P } _ { j } ( C I )  $. 
	\begin{flushright}
		$ \square $
	\end{flushright}
\end{exa}
	
\paragraph{Case 2: $ k $ is odd}
As for even $ k $, we initially assign probability mass $ \frac{n_{i}- 1}{n} $ to each category, $ {c}_{i} $, for $ i= 1, ..., k $. Then, as we aim to assign as much as possible of probability masses to the categories with the smallest $ \acute{f_{i}} $, we assign probability mass of $ \frac{2}{n} $ to each of $ {c}_{1} $ to $ {c}_{\frac{k}{2}-\frac{1}{2}} $. After that we assign the last remaining probability mass $ \frac{1}{n} $ to category $ {c}_{\frac{k}{2}+\frac{1}{2}} $.\\

\begin{exa} \label{k5}
	Consider the data set described in Example \ref{k6}, excluding the last category $ c_{6} $ and its corresponding fraction $ f_{6} $. So we have 5 possible categories where $ n= 15 $. In this case, first, each category is assigned its lower probability $ \frac{n_{i} - 1}{n} $. Thus, we assign $ (0, \frac{1}{15}, \frac{2}{15}, \frac{3}{15}, \frac{4}{15}) $ to $ (c_{1}, c_{2}, c_{3}, c_{4}, c_{5}) $, respectively. Secondly, we assign $ (\frac{2}{15}, \frac{2}{15}) $ to  $ (c_{1}, c_{2}) $, respectively. Finally, we assign the last remaining probability mass  $ \frac{1}{15} $ to  $ c_{3} $. Therefore, the final assignments are  $ (\frac{2}{15}, \frac{3}{15}, \frac{3}{15}, \frac{3}{15}, \frac{4}{15}) $ to $ (c_{1}, c_{2}, c_{3}, c_{4}, c_{5}) $, respectively.
	\begin{flushright}
		$ \square $
	\end{flushright}
\end{exa}

Following the above method in Case 1 and Case 2 of assigning probability masses to the possible categories over the `optimal configuration', we get the NPI lower probability for $ CI $, $ \underline { P } _ { j } ( C I ) $.

\subsubsection{Upper probability}
To find the NPI upper probability for $ CI $, we need to maximise $ p_{i} $ over all possible configurations on probability wheel, then to choose the configuration that gives the highest possible value of $ \overline { P } _ { j } ( C I ) $.\\

Let $ f_{i} = \displaystyle \frac { n ^ { C_{i} } ( t _ { j } = c_{i} ) + 1 } { n ( t _ { j } = c_{i} ) + 1 } $, for $ i= 1, ..., k $, so we rewrite the Formula (\ref{u-ci}) in the following way 
\begin{equation}\label{}
	\overline { P } _ { j } ( C I ) = \max\limits_{p_{i} \in \mathcal{P}} (f_{1} p_{1} + f_{2} p_{2} + ... + f_{k} p_{k})
\end{equation}
then rearranging $ f_{i} $ in an increasing order such that 
$ \acute{f}_{1} \leq \acute{f}_{2} \leq ... \leq \acute{f}_{k} $. So we get 
\begin{equation}\label{}
	\overline { P } _ { j } ( C I ) = \max\limits_{\acute{p}_{i} \in \mathcal{P}} (\acute{f}_{1} \acute{p}_{1} + \acute{f}_{2} \acute{p}_{2} + ... + \acute{f}_{k} \acute{p}_{k})
\end{equation} 

To maximise over $ \acute{p}_{i} $, we separate the largest categories as much as possible to ensure that we can assign probability masses of slices `in between' to its neighbour with larger value of $ \acute{f_{i}} $. Therefore, we consider the same configuration of the probability wheel (see Figure \ref{opt-conf}) that is used to find the NPI lower probability, but we want to assign as much as possible of probability masses to the categories with the largest $ \acute{f_{i}} $. Of course, each category is assigned its lower probability $ \frac{n_{i}- 1}{n}  $, and the remaining probability masses is shared between other categories in such way we get the largest value possible of $ CI $. We consider the following two cases to explain how such maximisation can be done.

\paragraph{Case 1: $ k $ is even}
As a first step, we assign probability mass $ \frac{n_{i}- 1}{n} $ to each category. As a second step, the remaining probability masses are then shared equally between the categories with the largest possible $ \acute{f_{i}} $. Hence, we assign probability mass of $ \frac{2}{n} $ to $ {c}_{\frac{k}{2}+1} $ to $ {c}_{k} $. This is clearly optimal to maximise the NPI upper probability for $ CI $, since we assign the same number of possible probability masses between the largest categories. Note that we can not assign more than probability mass of $ \frac{2}{n} $ to any category.\\

\begin{exa}
	Consider the same data described in Example \ref{k6}, where $ k = 6 $. In order to find the NPI upper probability for $ CI $, we assign $ (0, \frac{1}{21}, \frac{2}{21}, \frac{3}{21}, \frac{4}{21}, \frac{5}{21}) $ to $ (c_{1}, c_{2}, c_{3}, c_{4}, c_{5}, c_{6}) $, respectively. Then we assign $ (\frac{2}{21}, \frac{2}{21}, \frac{2}{21}) $ to  $ (c_{4}, c_{5}, c_{6}) $. Therefore, the final assignments are $ (0, \frac{1}{21}, \frac{2}{21}, \frac{5}{21}, \frac{6}{21}, \frac{7}{21}) $ to $ (c_{1}, c_{2}, c_{3}, c_{4}, c_{5}, c_{6}) $, respectively. Following theses assignments we get $ \overline { P } _ { j } ( C I )  $.
	\begin{flushright}
		$ \square $
	\end{flushright}
\end{exa}

\paragraph{Case 2: $ k $ is odd}
We initially assign probability mass $ \frac{n_{i}- 1}{n} $ to each category, $ {c}_{i} $, for $ i= 1, ..., k $. Then, the best way to distribute the remaining probability masses is to assign probability mass of $ \frac{2}{n} $ to $ {c}_{\frac{k}{2}+\frac{3}{2}} $ to $ {c}_{k} $. After that, to assign the last probability mass $ \frac{1}{n} $ to $ {c}_{\frac{k}{2}+\frac{1}{2}} $. Following this distributions, we assure that we get the largest value possible of the NPI upper probability for $ CI $\\

\begin{exa}
	Consider the data in Example \ref{k5}, where $ k = 5 $. To find the NPI upper probability for $ CI $, we assign $ (0, \frac{1}{15}, \frac{2}{15}, \frac{3}{15}, \frac{4}{15}) $ to $ (c_{1}, c_{2}, c_{3}, c_{4}, c_{5}) $, respectively. Then we further assign $ (\frac{2}{15}, \frac{2}{15}) $ to  $ (c_{4}, c_{5}) $, respectively, and $ \frac{1}{15} $ to  $ c_{3} $. Therefore, the final assignments are $ (0, \frac{1}{15}, \frac{3}{15}, \frac{5}{15}, \frac{6}{15}) $ to $ (c_{1}, c_{2}, c_{3}, c_{4}, c_{5}) $, respectively.
	\begin{flushright}
	    $ \square $
	\end{flushright}
\end{exa}

Following the above way of assigning probability masses to the possible categories over the `optimal configuration', we get the NPI upper probability for $ CI $, $ \overline { P } _ { j } ( C I ) $. Finally, we can use these lower and upper probabilities of $ CI $ to build classification trees using the D-NPI algorithm as explained in Section \ref{D-NPI algorithm}.\\

%%%%%%%%%%%%%%%%%%%%%%%%%%%%%%%%%%%%%%%%
%%%%%%%%%%%%%%%%%%%%%%%%%%%%%%%%%%%%%%%%
\section{The D-NPI classification tree algorithm }\label{D-NPI algorithm}
In this section we propose a new algorithm for classification trees which we call \textit{Direct Nonparametric Predictive Inference classification} (D-NPI) algorithm. The building process is similar to the well-known C4.5 algorithm (see Section \ref{backgr.}), but we use the $ CI $ method introduced in Section \ref{CI} as a split criterion to choose the best splitting attribute at each node. The $ CI $ split criterion is introduced for binary dataset in Section \ref{CI:B} and multinomial dataset in Section \ref{CI:M}.\\

Suppose that we have a training dataset, $ \mathcal{D} $, which has binary or categorical attribute variables. Let $ C $ be the target variable which also has binary or categorical possible results. For the training dataset, $ \mathcal{D} $, we first calculate the $ CI $ intervals, i.e. the lower and upper probabilities of $ CI $, for the complete list of attribute variables, $ t_{j} $, for $ j= 1,..., T $ using the lower and upper probabilities for $ CI $, see Equations (\ref{LCI}) and (\ref{UCI}) for binary data, and Equations (\ref{l-ci}) and (\ref{u-ci}) for multinomial data. Then, these results are compared with the lower and upper probabilities for no further attribute. We use the abbreviation $ \underline{P}({no\; att.}) $ and $ \overline{P}({no\; att.}) $ to indicate the lower and upper probabilities for no attribute, respectively. The lower and upper probabilities for no attribute correspond to simply stating the most common result in the target variable. Therefore, for binary data and using the NPI method for Bernoulli quantities \cite{Coolen_low.str}, introduced in Section \ref{backgr.}, the lower and upper probabilities for no attribute are 

\begin{equation}
	[\underline {P},\overline {P}](C = 1 | (n, s))  = \left[ \frac {s } { n+1 },  \frac {s+1} {n + 1 }\right]
\end{equation}\\
where $ s $ is the number of successes in $ n $ instances, or the most common results in the target variable in the case of having the number of failures is greater than the number of successes. For example, suppose we have a class variable with $ 70 $ instances in class state $ 0 $, and $ 30 $ instances in class state $ 1 $, where all instances equals to $ 100 $, then $ \underline{P}({no\; att.}) = \frac{s}{n+1} = \frac{70} {101} $, and $  \overline{P}({no\; att.}) =  \frac{s+1} {n + 1} = \frac{71}{101} $. For the case of multinomial data, the lower and upper probabilities for no attribute are simply the NPI-M lower and upper probabilities of the singleton event given in Equations (\ref{s-npi-m-l}) and (\ref{s-npi-m-u}) that correspond to the largest class in the target variable.\\

To find the best split attribute for each node, the attribute with the highest value of the lower and upper probabilities of $ CI $ is chosen. However, the lower probability of $ CI $ has to be greater than the lower probability for no attribute, and the upper probability of $ CI $ has to be greater than the upper probability for no attribute. In other words, there are two conditions in order to split upon the attribute variable with highest $ CI $ values, which are

\begin{equation}\label{2conditions}
	\underline{P}({CI}) > \underline{P}({no\; att.}) \;\;\;\; \mbox{and} \;\;\;\; \overline{P}({CI}) > \overline{P}({no\; att.})
\end{equation}\\
where the $ \underline{P}({CI})$ and $\overline{P}({CI}) $ correspond to the attribute variable with the highest lower and upper probabilities of $ CI $, and $ \underline{P}({no\; att.})$ and $\overline{P}({no\; att.}) $ correspond to the highest class of the target variable. The presence of the two conditions in Inequalities (\ref{2conditions}) is more likely to prevent overfitting in a classification tree generated by the D-NPI algorithm, as these conditions prevent us from building larger trees that may over fit the data and has less classification accuracy on testing set. It is noticed that during the experimental analysis, splitting attributes without considering these conditions does not improve the classification accuracy. Therefore, these two conditions are used as a stop criterion. However, it remains an open research question to study this stop criterion in more details. For example, using the imprecision along with these conditions particularly when there is overlap between two or more attributes that all fulfil these conditions or in other cases when one condition is met but not the another one. If there is not any attribute variable that fulfils the two conditions in Inequalities (\ref{2conditions}), we do not split further and transform the node into a leaf with the most common class in the target variable.\\

After selecting the best attribute variable $ t_{j} $ at the root node, we split the training dataset, $ \mathcal{D} $, into disjoint subsets $ \mathcal{D}_{i} $, where $ \mathcal{D}_{i} $ includes all instances with result $ t_{j}= c_{i} $, for $ i= 1, ..., k $ for the selected attribute variable. For binary data, $ \mathcal{D} $ is split into two disjoint subsets $ \mathcal{D}_{0} $ and $ \mathcal{D}_{1} $, where $ \mathcal{D}_{0} \bigcup \mathcal D_{1} = \mathcal{D} $ and $ \mathcal{D}_{0} \bigcap \mathcal{D}_{1} = \emptyset $, where $ \mathcal{D}_{i} $, for $ i = 0, 1 $, indicates both results of the chosen attribute variable. While for multinomial data, $ \mathcal{D}_{i} $ may contain more than two subsets. For example, if we choose an attribute variable to split on with 3 categories namely Low, Medium and High, then we would have 3 subsets of training dataset as $ \mathcal{D}_{1} $,  $ \mathcal{D}_{2} $ and  $ \mathcal{D}_{3} $. Then, we calculate the lower and upper probabilities of $ CI $ for each subset and compare the results of each subset with the corresponding lower and upper probabilities for no attribute. The D-NPI algorithm continues recursively by splitting further and hence, constructing new subtrees to each branch. The algorithm terminates when the two conditions in Inequalities (\ref{2conditions}) are not fulfilled or when the observations in the subset all belong to the same class, hence this class is used as a label to that corresponding leaf node. This process can be represented as a classification tree. Algorithm 1 describes the D-NPI algorithm. Note that this algorithm can be used to build classification trees for binary data or multinomial data, but the split criterion ($ CI $) is differently calculated for each case.\\

\begin{algorithm}
\caption{Pseudocode of the D-NPI algorithm.}
\footnotesize
 \begin{algorithmic}[1]
    \State \textbf{Input:}
  	\State \texttt{TR}: Training dataset
	\State \texttt{Target}: Target variable 
	\State \texttt{Attr}: List of attribute variables 
	\Procedure{D-NPI} {\texttt{TR}, \texttt{Target}, \texttt{Attr}}\\
	Create a Root node for the tree
	\If{\texttt{TR} have the same class C,}
	\State Return the single-node tree with class C
	\EndIf
	\If{Attr is empty,}
	\State Return the single-node tree with the most common class C in \texttt{TR}
	\EndIf 
	\State Otherwise
	\For{each attribute, $ t $ in \texttt{Attr}}
	\State Compute $\underline{P}(CI)$ and $\overline{P}(CI)$	
	\If{$\underline{P}(CI) > \underline{P}({no \,\,att.})$ and $\overline{P}(CI) > \overline{P}({no \,\,att.})$}
	\State Choose attribute, $ t $ with highest $\underline{P}(CI)$ and $\overline{P}(CI)$ 
	\Else 
	\State Add a leaf node labeled with the most common class in \texttt{TR} 
	\EndIf
	\EndFor
	\State Set $t$ the attribute for Root 
	\For{each value of $ t $, $ vi $,}
	\State Add a branch below Root, corresponding to $t= v_{i}$
	\State Let \texttt{TR}$_{v_{i}}$ be the subset of \texttt{TR} that have $t= v_{i}$
	\If{\texttt{TR}$_{v_{i}}$  is empty,}
	\State Add a leaf node labeled with the most common class in \texttt{TR}
	\Else 
	\State Add the subset generated by D-NPI(\texttt{TR}$_{v_{i}}$, \texttt{Target}, \texttt{Attr}-\{$t$\})
	\EndIf
	\EndFor\\
	\Return Root
	\EndProcedure
 \end{algorithmic}
\end{algorithm}

%%%%%%%%%%%%%%%%%%%%%%%%%%%%%%%%%%%%%%%%
%%%%%%%%%%%%%%%%%%%%%%%%%%%%%%%%%%%%%%%%
\section{Experimental Analysis}\label{exp. ana}
In this section, we examine the performance of the D-NPI algorithm on 13 datasets extracted from the UCI repository of machine learning databases \cite{uci}. The aim of this experimental analysis is not only to assess the performance of the D-NPI algorithm, but also to compare it with the most commonly used classical algorithm, the C4.5 algorithm, and with some algorithms based on imprecise probabilities such as the NPI-M, the A-NPI-M and the IDM (with two choices of the parameter $ s $). These algorithms are introduced in Section \ref{backgr.}.\\

%%%%%%%%%%%%%%%%%%%%%%%%%%%%%%%%%%%%%%%%
\subsection{Experimental setup}
The experiments were conducted using the R software. The datasets used in this experimental analysis are diverse in terms of their size, the number of classes and the range of categories of the attribute variables. A summary of the main characteristics of each data set is given in Table \ref{desc-multdata}. Further information on these datasets can be found in \cite{uci}. As a first step of developing the D-NPI algorithm, we apply the algorithm on only binary datasets, where both the class variable and attributes are binary. The first five datasets in Table \ref{desc-multdata} are used for this analysis. As we consider only binary datasets, all continuous attribute variables are converted to binary ones using the same thresholds given by Information Gain Ratio criterion \cite{quinlan93}. After that all classification tree algorithms are then built on these datasets. Note that the Acute Inflammations dataset has two target variables, we construct two trees based on each one. Thus, it is considered as two separate datasets. To construct the D-NPI algorithm on the first five datasets in Table \ref{desc-multdata}, we use the $ CI $ split criterion presented in Section \ref{CI:B}. On the other hand, the $ CI $ split criterion for multinomial datasets that is presented in Section \ref{CI:M} is used for the rest of the datasets.\\ 

The Nursery dataset is large, to reduce the amount of computation required in this data, we fix a minimum split number of 100 observations that must exist in order to split any node further or otherwise we terminate the tree and fix a leaf node with the most common class in that node. A minimum split value is sometimes fixed for a specific value to reduce the required computation as done by Bertsimas and Dunn \cite{Dunn-Berts}. However, this minimum split value is also used for all other algorithms applied to the Nursery dataset, hence the comparison will be fair although we might notice that the accuracy is lower than what is expected in this dataset when the minimum split value is not fixed. In the modified Iris dataset, we convert four continuous attributes to categorical attributes with three categories coded as, `L', `M' and `H' using equal frequency from the \texttt{`arules'} package in R and using \texttt{`discretize'} function. This function converts a continuous variable into a categorical variable using different strategies. The default one is used which is equal frequency. Note that the frequencies may not be exactly equal because of ties in the data. However, we aimed to convert all attributes to categorical ones, then to apply all classification algorithms on the converted data. This is because of that the D-NPI algorithm utilizes only categorical attributes. However, it will be of interest to generalise the D-NPI algorithm to deal with continuous attributes. All other datasets only have categorical attributes. All missing values were replaced with modal values. Finally, all classification algorithms have been applied to all these datasets under the same circumstances of pre-analysis steps to be a fair comparison.\\  

\begin{table}[t]
	\centering
	\footnotesize 
		\begin{tabular}{l c c c c}
			\hline
			Dataset & N & Att & Range of Att & Classes \\ 
			\hline 
			Acute Inflammations 1 &  120  & 6 & 2 & 2\\
			Acute Inflammations 2 &  120  & 6 & 2 & 2\\
			Banknote authentication & 1372 & 4 & 2 & 2\\
			Breast Cancer Wisconsin & 699 & 9 & 2 & 2\\
			Congressional Voting Records & 435 & 16 & 2 & 2 \\
			CMC & 1473 & 9 & 2-4 & 3 \\
			Hayes-Roth & 160 & 5 & 3-4 & 3 \\
			Lenses & 24 & 4& 2-3& 3\\
			Modified Iris & 150 & 4 & 3 &3\\
			Monk's Problems-1 & 124 & 7 & 2-4 &2\\
			Nursery & 12960 & 8 & 2-5 & 5\\
			Post-Operative Patient  & 90 & 8 & 2-4 & 3\\
			Qualitative-Bankruptcy & 250 & 6 & 3 & 2\\
			\hline 
		\end{tabular}
	\caption{Datasets description.}
	\label{desc-multdata}
\end{table}

Six classification algorithms have been used in this analysis, which are the D-NPI, the C4.5, the NPI-M, the  A-NPI-M and the IDM with $ s = 1 $ and $ s = 2 $. We denote the IDM with $ s = 1 $ and $ s = 2 $ by IDM1 and IDM2, respectively (see Section \ref{backgr.}). A 10-fold cross-validation procedure has been used for each dataset, then the average results are reported. Classification accuracy rates are used to measure and compare the performance of each classifier. It is the most commonly used method to measure the performance of classification algorithms. It is calculated as the ratio of the total number of correctly classified instances on the testing set to the total number of instances in the testing set. Given a sample confusion matrix as in Table \ref{conf.matrix}, the classification accuracy is 
\begin{center}
	$ \mbox{accuracy} =\frac{\mbox{TP + TN}}{\mbox{N}} $,
\end{center} 
where N = TN + TP + FN + FP, and TN, TP, FN and FP denote true negatives, true positives, false negatives and false positives, respectively.
\begin{table}[t]
	\centering  
	\footnotesize 
	\begin{tabular}{lcc}
		\hline
		& Class 1 (Predicted) & Class 2 (Predicted) \\ 
		\hline 
		Class 1 (Actual) & TN  & FP  \\
		Class 2 (Actual) & FN  & TP  \\
		\hline 
	\end{tabular}
	\caption{A sample confusion matrix.}
	\label{conf.matrix}
\end{table}
For further analysis of the D-NPI algorithm and to compare it with other classification algorithms, we used in-sample accuracy which is the classification accuracy rate on the training set. The classification accuracy and in-sample accuracy of each algorithm in each dataset are obtained from 10 runs using 10-fold cross validation, then the average results of these runs are reported as a final result.\\

%%%%%%%%%%%%%%%%%%%%%%%%%%%%%%%%%%%%%%%%
\subsection{Results}
First, the performance of the D-NPI classification algorithm has been evaluated against five other classification algorithms. Table \ref{acc-results} presents the classification accuracies of the proposed D-NPI classification algorithm and all other algorithms for each dataset. The results in Table \ref{acc-results} indicate that the D-NPI classification algorithm slightly outperforms other classification algorithms in 9 datasets. The Acute Inflammations dataset has been created by a medical expert as a dataset to test an expert system, which will perform the presumptive diagnosis of two diseases of the urinary system \cite{uci}. All algorithms have achieved the full classification accuracy on this dataset either Acute Inflammations 1 or Acute Inflammations 2, except the D-NPI algorithm in Acute Inflammations 1. However, the D-NPI algorithm produces relatively smaller trees than the other algorithms for this dataset, where the tree size is considered as the total number of leaves. For the Banknote authentication, Breast Cancer Wisconsin and Congressional Voting Records datasets, the D-NPI algorithm performs slightly better than the other algorithms, although all algorithms have a very similar classification accuracies.\\

For the Lenses dataset, it is noticed that there is a clear difference in the classification accuracies among classification algorithms, where the D-NPI classification algorithm clearly outperforms the C4.5 and the IDM1, and slightly superior to the NPI-M, the A-NPI and the IDM2. This clear difference in the classification accuracy could be because it is a small dataset with only 24 observations. For the Monk's Problem-1 dataset, the D-NPI classification algorithm has the highest classification accuracy of 73.33\%, where all other algorithms have the same accuracy of 69.17\%. For this dataset, the D-NPI classification algorithm returns relatively larger trees than other algorithms which might be the reason of its superiority. For the Post-Operative Patient dataset, the NPI-M, the A-NPI-M and the IDM classification algorithms have the highest classification accuracy rates. For the Hayes-Roth dataset, we can see that the IDM2 is slightly superior to all other algorithms with classification accuracy of 67.69\%. For the CMC, Modified Iris, Nursery and Qualitative-Bankruptcy datasets, the D-NPI classification algorithm is superior to all the other algorithms. Overall, according to the average classification accuracy rate, we can say that all classification algorithms are performing similarly, but with a slightly better performance of the D-NPI algorithm.\\  

\begin{table}[t]
	\centering
	\footnotesize 
	\begin{tabular}{lcccccc}
		\hline
		Dataset & D-NPI & C4.5 & NPI-M & A-NPI-M & IDM1 & IDM2 \\ 
		\hline 
		Acute Inflammations 1 & 94.17 & \textbf{100} &  \textbf{100} & \textbf{100} & \textbf{100} & \textbf{100} \\
		Acute Inflammations 2 & \textbf{100} & \textbf{100} & \textbf{100} & \textbf{100} & \textbf{100} & \textbf{100} \\
		Banknote authentication & \textbf{89.50} & 89.49 & 89.49 & 89.49 & 89.49 & 89.49 \\
		Breast Cancer Wisconsin & \textbf{94.85} & 94.27 & 93.19 & 93.19 & 93.19 & 93.19 \\
		Congressional Voting Records  & \textbf{95.64} & 95.58 & 95.58 & 95.58 & 95.58 & 95.35\\
		CMC & \textbf{45.49} & 45.31 & 42.93 & 42.93 & 42.93 & 42.93 \\
		Hayes-Roth & 66.76 & 66.92 & 64.62 & 64.62 & 63.08 & \textbf{67.69}\\
		Lenses &  \textbf{81.67} & 70.00 & 80.00 & 80.00 & 75.00  & 80.00\\
		Modified Iris &  \textbf{95.33} & 92.00 & 90.00 & 90.00 & 92.67  & 92.67\\
		Monk's Problems-1 & \textbf{73.33} & 69.17 & 69.17 & 69.17 & 69.17 & 69.17\\
		Nursery & \textbf{90.37} & 89.21 & 89.20 & 89.20 & 89.20  & 89.20\\
		Post-Operative Patient & 67.78 & 68.89 & \textbf{71.11} & \textbf{71.11} & \textbf{71.11} &  \textbf{71.11} \\
		Qualitative-Bankruptcy & \textbf{99.60} & 98.00 & 98.40 & 98.40 & 98.40 & 98.40 \\
		Average & \textbf{84.19} & 82.99 & 83.36 & 83.36 & 83.07 & 83.79 \\
		\hline 
	\end{tabular}
	\caption{Classification accuracy results for classification algorithms built with 10-fold cross validation.}
	\label{acc-results}
\end{table}

Secondly, following \cite{Dunn-Berts,Murthy-Salzberg}, we have used in-sample accuracy rate to measure the performance of the D-NPI classification algorithm on the training set, and to compare it with the other classification algorithms. The in-sample accuracy measure is not commonly used to indicate classification accuracy, but it gives insight into how the algorithm performs on the training set. It is known that if the classification algorithm performs very well on the training set but not very well on the testing set, this is likely to indicate overfitting. Thus, the in-sample accuracy is reported to show the performance of classification algorithms on both training and testing sets. Table \ref{inacc-results} shows the in-sample accuracy results of classification algorithms built with 10-fold cross validation. The D-NPI classification algorithm slightly outperforms the other classification algorithms in several datasets, followed by the C4.5 classification algorithm which is also superior to other algorithms in some datasets. It is noticed that the IDM2 algorithm slightly outperforms the IDM1 algorithm with regards to both average classification accuracy and average in-sample accuracy rates. In this experimental analysis, the NPI-M and the A-NPI-M algorithms are equivalent in all performance measures (i.e.\ classification accuracy, in-sample accuracy and tree size). Finally, the D-NPI classification algorithm has the highest average result of in-sample accuracy compared to the other classification algorithms. It should be clarified that the D-NPI algorithm has good results on in-sample accuracy and classification accuracy as well, which may indicate that it does not suffer from overfitting.\\    

\begin{table}[t]
	\centering
	\footnotesize 
	\begin{tabular}{l c c c c c c}
		\hline
		Dataset & D-NPI & C4.5 & NPI-M & A-NPI-M & IDM1& IDM2\\ 
		\hline
		Acute Inflammations 1 & 94.17 & \textbf{100} & \textbf{100} & \textbf{100} & \textbf{100} & \textbf{100} \\
		Acute Inflammations 2 & \textbf{100} & \textbf{100} & \textbf{100} & \textbf{100} & \textbf{100} & \textbf{100} \\
		Banknote authentication & \textbf{89.51} & \textbf{89.51} & \textbf{89.51} & \textbf{89.51} & \textbf{89.51} & \textbf{89.51} \\
		Breast Cancer Wisconsin & \textbf{95.31} & 94.20 & 93.67 & 93.67 & 93.84 & 93.67 \\
		Congressional Voting Records  & 95.63 & \textbf{95.64} & \textbf{95.64} & \textbf{95.64} & \textbf{95.64} & \textbf{95.64}\\ 
		CMC & \textbf{48.84} & 47.22 & 45.17 & 45.17 & 45.17 & 45.17 \\
		Hayes-Roth & \textbf{83.67} & 81.51 & 79.33 & 79.33 & 81.51 & 82.35 \\
		Lenses &  \textbf{87.49} & 84.55 & 85.91 & 85.91 & 85.00 & 85.91 \\
		Modified Iris &  95.33 & \textbf{95.41}  & 92.07 & 92.07 & 94.07 & 94.07\\
		Monk's Problems-1 & \textbf{84.24} & 74.82  & 73.66 & 73.66 & 73.66 & 73.66 \\
		Nursery & \textbf{90.37} & 89.21 & 89.26 & 89.26 & 89.26 & 89.26 \\
		Post-Operative Patient & 71.23 & \textbf{71.36} & 71.11 & 71.11 & 71.11 & 71.11 \\
		Qualitative-Bankruptcy & \textbf{99.60} & 99.24 & 98.40 & 98.40 & 98.40 & 98.40 \\
		Average & \textbf{87.34} & 86.36 & 85.67 & 85.67 & 85.93 & 86.06 \\
		\hline 
	\end{tabular}
	\caption{In-sample accuracy results for classification algorithms built with 10-fold cross validation.}
	\label{inacc-results}
\end{table}

Thirdly, in order to compare different trees generated by the classification algorithms, an average tree size of each algorithm is reported. Table \ref{average-tree size} shows the average results of tree size for each classification algorithm. Note that we refer to tree size as the total number of leaf nodes as was done by Bertsimas and Dunn \cite{Dunn-Berts}, and Murthy and Salzberg \cite{Murthy-Salzberg}. However, other researchers may consider the total number of all nodes. Of course both methods can be used to refer to the tree size. It can be observed from Table \ref{average-tree size} that the average tree size of all algorithms is nearly equivalent with some smaller trees generated by the C4.5 algorithm followed by the D-NPI algorithm. From the experimental analysis that is done on all these datasets, it is noticed that the D-NPI algorithm generates relatively smaller trees than other algorithms when applied to binary datasets, but it does not have the smallest trees with regards to multinomial datasets.\\

To summarise, from the classification accuracy given in Table \ref{acc-results}, we draw the following conclusions about the performance of the D-NPI algorithm. The D-NPI algorithm is performing well and slightly outperforms other algorithms. On the other hand, the NPI-M and the A-NPI-M are performing the same on all these datasets. The IDM2 outperforms the IDM1 algorithm with regards to this measure. With regards to the in-sample accuracy, the D-NPI algorithm is slightly superior to other algorithms followed by the C4.5 algorithm. Finally, the C4.5 algorithm has the smallest average tree size, while the IDM1 algorithm has the largest average tree size.\\

\begin{table}[t]
	\centering
	\footnotesize 
	\begin{tabular}{lcccccc}
		\hline
		Algorithm & D-NPI & C4.5 & NPI-M & A-NPI-M & IDM1 & IDM2 \\ 
		\hline
		Average & 6.48 & \textbf{5.98} & 7.17 & 7.17 & 7.64 & 6.93 \\
		\hline 
	\end{tabular}
	\caption{Average tree size for classification algorithms built with 10-fold cross validation.}
	\label{average-tree size}
\end{table} 

%%%%%%%%%%%%%%%%%%%%%%%%%%%%%%%%%%%%%%%%
%%%%%%%%%%%%%%%%%%%%%%%%%%%%%%%%%%%%%%%%
\section{Conclusions and future works}\label{concl.}
In this paper, we have proposed a new algorithm to build classification trees from Nonparametric Predictive Inference perspective, which we call the Direct Nonparametric Predictive Inference (D-NPI) algorithm. The D-NPI classification algorithm uses a new split criterion, Correct Indication ($ CI $), which is based on the lower and upper probabilities for NPI for binary and multinomial data. The performance of the D-NPI classification algorithm has been tested against different classification tree algorithms using different performance measures on many datasets from the UCI repository of machine learning. It has been shown that the D-NPI classification algorithm slightly performs better than the other classification algorithms.\\

In future works, it would be of interest to extend this research work further to involve real-valued data. It would be also interesting to explore the use of the D-NPI classification algorithm in random forests. Another interesting extension to this work is to develop the D-NPI algorithm with imprecise classification, which might return a set of states rather than the most single state in the class variable. Another work which is currently under investigation is measuring the performance of the D-NPI algorithm when it is applied to noisy data. Finally, exploring the use of imprecision as a stop criterion may lead to improvements on the D-NPI classification algorithm.\\

\section*{Acknowledgements}
Abdulmajeed Alharbi gratefully acknowledges the financial support received from Taibah University in Saudi Arabia and the Saudi Arabian Cultural Bureau (SACB) in London for pursuing his PhD studies at Durham University.
\newpage
%%%%%%%%%%%%%%%%%%%%%%%%%%%%%%%%%%%%%%%%%%%%%%%%%%%%%%%%%%%%%%%%%%%%%%%%%%%%%%%%%%%%%%%

\end{document}